\documentclass{article}
\usepackage{amsmath}
\usepackage{graphics}

\input{tcilatex}
\begin{document}

\centerline{\Large \bf Convergence properties of  fixed-point
search} \centerline{\Large \bf  with general but equal phase shifts
for any number of iterations}

\footnote{%
The paper was supported by NSFC(Grants No. 60433050 and 60673034), the basic
research fund of Tsinghua university NO: JC2003043.}

\centerline{Dafa Li$^{a}$\footnote{email
address:dli@math.tsinghua.edu.cn},  Xiangrong Li$^{b}$, Hongtao
Huang$^{c}$, Xinxin Li$^{d}$ }

\centerline{$^a$ Department of mathematical sciences, Tsinghua
University, Beijing 100084 CHINA}

\centerline{$^b$ Department of Mathematics, University of
California, Irvine, CA 92697-3875, USA}

\centerline{$^c$ Electrical Engineering and Computer Science Department} %
\centerline{ University of Michigan, Ann Arbor, MI 48109, USA}

\centerline{$^d$ Department of computer science, Wayne State
University, Detroit, MI 48202, USA}

The correspondence Author Dafa Li,

Phone Number is (8610)62773561

Fax No. is (8610) 62785847

Abstract

Grover presented the fixed-point search by replacing the selective
inversions by\ selective phase shifts of $\pi /3$. In this paper, we
investigate the convergence behavior of the fixed-point search algorithm
with general but equal phase shifts for any number of iterations.




\section{Introduction}

Grover's quantum search algorithm is used to find a target state in an
unsorted database of size $N$\cite{Grover97}\cite{Grover98}. The Grover's
quantum search algorithm can be considered as a rotation of the state
vectors in two-dimensional Hilbert space generated by the start ($s$) and
target ($t$) vectors\cite{Grover98}. The amplitude of the target state
increases monotonically towards its maximum and decreases monotonically\
after reaching the maximum \cite{LDF05}. This search algorithm is called the
amplitude amplification algorithm. For the size $N=2^{n}$ of the database,
quantum search algorithm requires $O(\sqrt{N})$ steps to find the target
state.\ As mentioned in \cite{Grover05} \cite{Brassard97}, unless we stop
when it is right at the target state, it will drift away. A fixed-point
search algorithm was presented in \cite{Grover05} to avoid drifting away
from the target state. The fixed-point search algorithm obtained by
replacing the selective inversions by\ selective phase shifts of $\pi /3$,
converges to the target state irrespective of the number of iterations. The
main advantage of the fixed-point search with equal phase shifts of $\pi /3$
is that it performs well for small but unknown initial error probability and
the fixed-point behavior leads to robust quantum search algorithms \cite%
{Grover05}. However, the target state is the limit state when the number of
iterations tends to the infinite.

For readability, we introduce the fixed-point search algorithm as follows.
In \cite{Grover05} the transformation $UR_{s}^{\pi /3}U^{+}R_{t}^{\pi /3}U$,
where $U$ is any unitary operator, was applied to the start state $|s\rangle
$,

\begin{eqnarray}
R_{s}^{\pi /3} &=&I-[1-e^{i\frac{\pi }{3}}]|s\rangle \langle s|,  \notag \\
R_{t}^{\pi /3} &=&I-[1-e^{i\frac{\pi }{3}}]|t\rangle \langle t|,
\label{grover1}
\end{eqnarray}

\noindent where $|t\rangle $ stands for the target state. The transformation
$UR_{s}^{\pi /3}U^{+}R_{t}^{\pi /3}U$ is denoted as Grover's the Phase-$\pi
/3$ \ search algorithm in \cite{Tulsi}.

Let us consider the fixed-point search algorithm with general but equal
phase shifts as follows.

\begin{eqnarray}
R_{s}^{\theta } &=&I-[1-e^{i\theta }]|s\rangle \langle s|,  \notag \\
R_{t}^{\theta } &=&I-[1-e^{i\theta }]|t\rangle \langle t|.  \label{grover2}
\end{eqnarray}

\noindent The transformation $UR_{s}^{\theta }U^{+}R_{t}^{\theta }U$ was
called as the Phase-$\theta $ search algorithm and studied in \cite{LDF07a}.
It is enough to let $\theta $ be in $[0,\pi ]$.

Note that if we apply $U$ to the start state $|s\rangle $, then the
amplitude of reaching the target state $|t\rangle $ is $U_{ts}$\cite%
{Grover98}, where $\left\vert \left\vert U_{ts}\right\vert \right\vert
^{2}=1-\epsilon $.\ As indicated in \cite{Grover98}, in the case of database
search, $|U_{ts}|$ is almost $1/\sqrt{N}$, where $N$ is the size of the
database. Thus, $\epsilon $ is almost $1-1/N$ and $\epsilon $ is close to $1$
for the large size of database.

Apply the operations $U$, $R_{s}^{\theta }$, $U^{+}$, $R_{t}^{\theta }$, and
$U$\ to the start $|s\rangle $ and\ let $D(\theta )$ be the deviation of the
state $UR_{s}^{\theta }U^{+}R_{t}^{\theta }U|s\rangle $\ from the $t$ state
for any phase shifts of $\theta $. The deviation $D(\theta )$ was reduced in
\cite{LDF07a} and is rewritten as follows.
\begin{equation}
D(\theta )=4(1-\cos \theta )^{2}\epsilon (\epsilon -d)^{2},  \label{dev1}
\end{equation}%
where $d=\frac{1-2\cos \theta }{2(1-\cos \theta )}$. It was shown that $%
D(\theta )$ is between $0$ and $1$ in \cite{LDF07a}. For the Phase-$\pi /3$
search algorithm, $D(\pi /3)=\epsilon ^{3}$\cite{Grover05}.

$\allowbreak $In \cite{LDF06}, we explored the performance of the
fixed-point search with general but different phase shifts for one
iteration. In \cite{LDF07a}, we discussed the performance of the fixed-point
search with general but equal phase shifts for one iteration.

In this paper, we investigate convergence behavior of the fixed-point search
with general but equal phase shifts\ for any number of iterations. It is
useful for designing fixed-point search algorithms for different choices of
the phase shift parameter $\theta $. The following results are established
in Section 2.

(1). The fixed-point search with equal phase shifts of $\theta \leq \pi /2$
converges to the target state.

(2). The fixed-point search with equal phase shifts of $\theta $, where $\pi
/2<\theta \leq \arccos (-1/4)$, converges the target state with the
probability of at least $80\%.$

(3). The fixed-point search with equal phase shifts of $\theta $, where $%
\arccos (-1/4)<\theta \leq 2\pi /3$, converges the target state with the
probability of among $66.6\%$ and $80\%.$

(4). The fixed-point search with equal phase shifts of $\theta $, where $%
2\pi /3<\theta \leq \pi $, does not converge.

In section 3, we analyze the convergence rate for different values of $%
\theta $. It is demonstrated that the Phase-$\pi /3$ is not always optimal
and the convergence rate can be improved by choosing $\theta >\pi /3$. In
section 4, we show that for the size $N=2^{n}$ of the database, $O(n)$
iterations of the Phase-$\theta $ search\ can find the target state.
However, as indicated in \cite{Grover05}, $O(n)$ iterations of the Phase-$%
\theta $ search involve the exponential queries.

\section{Convergence performance of the Phase-$\protect\theta $ search for
any number of iterations}

Let $\epsilon _{0}=\epsilon $ and $0<\epsilon <1$.\ Then, from Eq. (\ref%
{dev1}) one can obtain the following iteration equation

\begin{equation}
\epsilon _{m+1}=4(1-\cos \theta )^{2}\epsilon _{m}(\epsilon _{m}-d)^{2}.
\label{g-iteration}
\end{equation}%
In this section, we discuss the convergence behavior of the Phase-$\theta $
search for any number of iterations. For the Phase-$\pi /3$ search, after
recursive application of the basic iteration for $m$ times, the failure
probability $\epsilon _{m}=$ $\epsilon ^{3^{m}}$ and the success probability
$\left\vert U_{m,ts}\right\vert =1-\epsilon ^{3^{m}}$\cite{Grover05}. The $%
\epsilon _{m}$\ in Eq. (\ref{g-iteration})\ is the failure probability of
the Phase-$\theta $ search algorithm after $m$ iterations.

From\ inference [11] in \cite{LDF07a}, Eq. (\ref{g-iteration}) has the
following fixed-points: $0$, $1$ ($\theta \neq 0$), $a$, where $a=\cos
\theta /(\cos \theta -1)$ ($\theta \neq 0$). In other words, if the sequence
$\{\epsilon _{m}\}$ in Eq. (\ref{g-iteration}) has a limit then the limit
must be $0$, $1$ or $a$. Clearly $a<d$.\

To study the convergence performance for any number of iterations, we need
the following results which are listed in the following paragraphs (A), (B),
and (C).

(A). From Eq. (\ref{g-iteration}), we obtain the following,

\begin{equation}
\epsilon _{m}-\epsilon _{m-1}=4\epsilon _{m-1}\left( \cos \theta -1\right)
^{2}\left( 1-\epsilon _{m-1}\right) (a-\epsilon _{m-1}).  \label{iteration-2}
\end{equation}

Eqs. (\ref{g-iteration}) and (\ref{iteration-2}) imply the following
convergence property. \ \ \

Property 1.

(1.1) If $\epsilon _{m}=d$, then $\epsilon _{m+l}=0$, for any $l>0$.

(1.2). if $\epsilon _{m-1}>a$ and $\epsilon _{m-1}\neq 0$, $\epsilon
_{m}<\epsilon _{m-1}$;

(1.3). If $\epsilon _{m-1}<a$ and $\epsilon _{m-1}\neq 0$, $\epsilon
_{m}>\epsilon _{m-1}$.

(B). When $\pi /2\leq \theta \leq \pi $, we have the following equation.

\begin{equation}
\epsilon _{m}-a=4(\cos \theta -1)^{2}(\epsilon _{m-1}-a)(\epsilon
_{m-1}-b)(\epsilon _{m-1}-c),  \label{fp-eq1}
\end{equation}%
where $b=\frac{1}{2}-\frac{\sqrt{-\cos \theta (2-\cos \theta )}}{2(1-\cos
\theta )}$, and $c=\frac{1}{2}+\frac{\sqrt{-\cos \theta (2-\cos \theta )}}{%
2(1-\cos \theta )}$. When $\epsilon _{i}=a$, $b$ or $c$, $\epsilon _{i+l}=a$
for any $l>0$.\ Note that $b<d$, $a<d$, and $d<c$.

$\allowbreak $(C). Let

\begin{equation}
f(x)=4(1-\cos \theta )^{2}x(x-d)^{2}.  \label{iterat}
\end{equation}%
Then, the derivative of $f(x)$ is

\begin{equation}
f^{\prime }(x)=12(1-\cos \theta )^{2}(x-d)(x-d/3).  \label{deriv}
\end{equation}
From Eq. (\ref{deriv}), (1). $f^{\prime }(x)=0$ at $r$ and $d$, where $r=d/3$%
; (2).$\ f^{\prime }(x)<0$ when $r<x<d$; (3). $f^{\prime }(x)>0$ when $x<r$
or $x>d$; (4). When $\theta >\pi /3$, $f(x)$ has a relative maximum $g=\frac{%
2(1-2\cos \theta )^{3}}{27(1-\cos \theta )}$ at $r$ and a relative minimum $%
0 $\ at $d$.

\subsection{When $0<\protect\theta \leq \protect\pi /2$, for any $\protect%
\epsilon _{0}\in (0,1)$, the Phase-$\protect\theta $ search converges to the
target state.}

Note that $0$ is an attractive fixed-point when $0<\theta <\pi /2$\ and $0$
is also a semi-attractive fixed-point when $\theta =\pi /2$. See inference
[11] in \cite{LDF07a}.

(1). $0<\theta \leq \pi /3$

For this case, $d\leq 0$ and $a<0$. In Eq. (\ref{g-iteration}), $d=0$ means $%
\epsilon _{m+1}=\epsilon _{m}^{3}$, which is Grover's Phase-$\pi /3$ search.
From $d<0$ and Eq. (\ref{g-iteration}), $\epsilon _{m}>0$.\ By property
(1.2), always $\epsilon _{m}<\epsilon _{m-1}$ when $0<\theta \leq \pi /3$.
That is, the sequence $\{\epsilon _{m}\}$ in Eq. (\ref{g-iteration})
decreases monotonically$.$Therefore, for any $\epsilon _{0}$ in $(0$, $1)$ $%
\lim_{m\rightarrow \infty }\epsilon _{m}=0$.

(2). $\pi /3<\theta \leq \pi /2$

For this case, $a\leq 0$, $0<d\leq 1/2$. Hence, from Eq. (\ref{g-iteration})
$0\leq \epsilon _{i}<1$. By property (1.2), always $\epsilon _{m}\leq
\epsilon _{m-1}$ when $\pi /3<\theta \leq \pi /2$. That is, the sequence $%
\{\epsilon _{m}\}$ in Eq. (\ref{g-iteration}) decreases. Factually, the
sequence $\{\epsilon _{m}\}$ in Eq. (\ref{g-iteration}) decreases
monotonically and $\epsilon _{m}>0$, or is of the form $\epsilon
_{0}>\epsilon _{1}>...>\epsilon _{k}=0$ and $\epsilon _{l}=0$ for any $l>k$.
Therefore, for any $\epsilon _{0}$ in $(0$, $1)$ $\lim_{m\rightarrow \infty
}\epsilon _{m}=0$.

Example 1. For the Phase-$\pi /2$ search, $\epsilon _{m}=\epsilon
_{m-1}(2\epsilon _{m-1}-1)^{2}$. Let $\epsilon _{0}=0.99999$. See Fig. 1.

$\epsilon _{1}=\allowbreak 0.999\,95$, $\ \ \epsilon _{2}=\allowbreak
0.999\,75$, $\epsilon _{3}=\allowbreak 0.998\,75$, $\ \ \epsilon
_{4}=\allowbreak 0.993\,76$,

$\epsilon _{5}=\allowbreak 0.969\,11$, $\ \ \epsilon _{6}=\allowbreak
0.853\,07$, $\epsilon _{7}=\allowbreak 0.425\,37$, $\ \ \epsilon
_{8}=\allowbreak 9.\,\allowbreak 476\,6\times 10^{-3}$.

\subsection{When $\protect\pi /2<\protect\theta <\arccos (-1/4)$, the Phase-$%
\protect\theta $ search converges the target state with the success
probability of $(1-a)>80\%$.\ }

For the Phase-$\theta $ search, $a<g<r<b<d<c$. Note that $a$ is an
attractive fixed-point. See \ inference [11] in \cite{LDF07a}. From Eq. (\ref%
{fp-eq1}), we have the following property.

Property 2.

(2.1). $a<\epsilon _{m}\leq g$ whenever $a<\epsilon _{m-1}<b$;

(2.2). $0\leq \epsilon _{m}<a$ whenever $b<\epsilon _{m-1}<c$ or $\epsilon
_{m-1}<a$.

\textbf{The convergence region of the Phase-}$\theta $\textbf{\ search}

(A). When $\epsilon _{0}\in (0,c]$ and $\epsilon _{0}\neq d$, the deviation
from the target state converges to the fixed-point $a$.

There are four cases. The argument is the following.

Case 1. When $\epsilon _{0}=a$ or $b$ or $c$, it is trivial by Eq. (\ref%
{fp-eq1}).

Case 2. $\epsilon _{0}<a$. By property (2.2), $0<\epsilon _{m}<a$ for any $m$%
. By property 1, the sequence $\{\epsilon _{m}\}$ increases monotonically.
Hence, the sequence $\{\epsilon _{m}\}$ converges to $a$ from below.

Case 3. $a<\epsilon _{0}<b$. By property (2.1), always $a<\epsilon _{m}\leq
g $ for any $m>0$, and by property 1, the sequence $\{\epsilon _{m}\}$
decreases monotonically. Hence, the sequence $\{\epsilon _{m}\}$ converges
to $a$ from above.

Case 4. $b<\epsilon _{0}<c$ and $\epsilon _{0}\neq d$. By property (2.2), $%
0<\epsilon _{1}<a$. Then, it turns to case 2.

Conclusively, when $\epsilon _{0}\in (0,c]$ and $\epsilon _{0}\neq d$, from
the above four cases, $\epsilon _{m}\neq d$, hence $\lim_{m\rightarrow
\infty }\epsilon _{m}=a$.

(B). When $\epsilon _{0}\in (c,1)$, the deviation from the target state
converges to the fixed-points $a$ or $0$.

By property (1.2), $\epsilon _{0}>...>\epsilon _{j^{\ast }-1}(>c)>\epsilon
_{j^{\ast }\text{ }}(\leq c)$. If $\epsilon _{j^{\ast }\text{ }}=d$, then $%
\epsilon _{m}=0$ for any $m>j^{\ast }$. Otherwise, $\lim_{m\rightarrow
\infty }\epsilon _{m}=a$ by the above (A).

\subsection{Phase-$\arccos (-1/4)$ search converges the target state with
the success probability of $80\%$.\ }

For the Phase-$\arccos (-1/4)$ search, $a=1/5$ is an attractive fixed-point,
see inference [11] in \cite{LDF07a}. $b=a=1/5$, $d=3/5$, and $c=4/5$. The
iteration equation is $\epsilon _{m}=\epsilon _{m-1}(\allowbreak 5\epsilon
_{m-1}-3)^{2}/4$. Eq. (\ref{fp-eq1}) becomes the following.

$\allowbreak $%
\begin{equation}
\epsilon _{m}-1/5=\allowbreak \frac{25}{4}\left( \epsilon _{m-1}-4/5\right)
\left( \epsilon _{m-1}-1/5\right) ^{2}.  \label{eq2}
\end{equation}

From Eq. (\ref{eq2}) we have the following property.

Property 3.

(3.1). $\epsilon _{m}<1/5$ when $\epsilon _{m-1}<4/5$ and $\epsilon
_{m-1}\neq 1/5$.

(3.2).\ $\epsilon _{m}>1/5$ when $\epsilon _{m-1}>4/5$.

\textbf{The convergence region of the Phase-}$\arccos (-1/4)$\textbf{\ search%
}

(A). When $\epsilon _{0}\in (0,4/5]$ and $\epsilon _{0}\neq 3/5$, the
deviation from the target state converges to the fixed-point $1/5$.

When $\epsilon _{0}=1/5$ or $4/5$, it is trivial by Eq. (\ref{fp-eq1}). When
$\epsilon _{0}\in (0,4/5)$ and $\epsilon _{0}\neq 1/5$, always $\epsilon
_{m}<1/5$ for $m>0$\ by property (3.1) and the sequence $\{\epsilon _{m}\}$
increases monotonically from $m>0$ by property (1.3). Therefore, the
sequence $\{\epsilon _{m}\}$ converges to $1/5$ from below.\ \

(B) When $\epsilon _{0}\in (4/5,1)$, the deviation from the target state
converges to the fixed-points $1/5$ or $0$.

By property (1.2), $\epsilon _{0}>\epsilon _{1}>....>\epsilon _{m}(\leq 4/5)$%
. Case 1. If $\epsilon _{m}=3/5$, then $\epsilon _{i}=0$ for any $i>m$. Case
2. Otherwise, by the above (A), $\lim_{m\rightarrow \infty }\epsilon
_{m}=1/5 $.

$\allowbreak $

Example 2. Let $\epsilon _{0}=0.9999;$

$\epsilon _{1}=\allowbreak 0.999\,4$, $\ \ \epsilon _{2}=\allowbreak
0.996\,4 $, $\ \ \ \ \ \ \ \ \epsilon _{3}=\allowbreak 0.978\,55$, $\
\epsilon _{4}=\allowbreak 0.876\,41$,

$\epsilon _{5}=\allowbreak 0.418\,50$, $\ \epsilon _{6}=\allowbreak
8.\,\allowbreak 616\,5\times 10^{-2}$, $\ \epsilon _{7}=\allowbreak
0.142\,19 $, $\ \epsilon _{8}=\allowbreak 0.186\,26$,

$\epsilon _{9}=\allowbreak 0.199\,28$, $\ \epsilon _{10}=\allowbreak 0.2$.

\subsection{$\allowbreak $When $\arccos (-1/4)<\protect\theta \leq 2\protect%
\pi /3$, the Phase-$\protect\theta $ search converges the target state with
the success probability of $(1-a)$, where $66\%\leq (1-a)<80\%$.\ }

For the Phase-$\theta $ search, $b<r<a<g<d<c$. Note that $a$ is an
attractive fixed-point when $\arccos (-1/4)<\theta <2\pi /3$\ and $1/3$ is a
semi-attractive fixed-point when $\theta =2\pi /3$. See inference [11] in
\cite{LDF07a}. From Eq. (\ref{fp-eq1}), we have the following property.

Property 4.

(4.1). $a<\epsilon _{m}\leq g$ whenever $b<\epsilon _{m-1}<a$;

(4.2). $0\leq \epsilon _{m}<a$ whenever $a<\epsilon _{m-1}<c$ or $\epsilon
_{m-1}<b$.

\textbf{The convergence region of the Phase-}$\theta $\textbf{\ search}

(A). When $\epsilon _{0}\in (0,c]$ and $\epsilon _{0}\neq d$, the deviation
from the target state converges to the fixed-point $a$.

There are seven cases. We argue them as follows.

Case 1. If $\epsilon _{0}=a$ or $b$ or $c$, then it is trivial.

Case 2. $a<\epsilon _{0}\leq g$.\ The proof is put in Appendix A.

Case 3. $\epsilon _{0}<b$. By property (1.3), $\epsilon _{j}$ increases
monotonically from $\epsilon _{0}$ until $\epsilon _{j^{\ast }-1}<b$ and $%
b\leq \epsilon _{j^{\ast }}<f(b)=a$ since $f^{\prime }(x)>0$ when $x<b$. If $%
\epsilon _{j^{\ast }}=b$, it is trivial. Otherwise, by property (4.1), $%
\epsilon _{j^{\ast }+1}$ is in $(a,g]$. Now it turns to case 2.

Case 4. $b<\epsilon _{0}\leq r$. When $\epsilon _{0}=r$, $\epsilon
_{m}=f^{(m-1)}(g)$. From the proof of case 2, $\lim_{m\rightarrow \infty
}f^{(m-1)}(g)=a$. Next consider that $b<\epsilon _{0}<r$. Since $f^{\prime
}(x)>0$ when $b<x<r$, $f(b)<f(\epsilon _{0})<f(r)$. That is, $a<\epsilon
_{1}<g$. It turns to case 2.

Case 5. $r<\epsilon _{0}<a$. Since $f^{\prime }(x)<0$ when $r<x<a$, $%
a<\epsilon _{1}<g$. It turns to case 2.

Case 6. $g<\epsilon _{0}<d$. Since $f^{\prime }(x)<0$ when $g<x<d$ and $a<g$%
, $0<\epsilon _{1}<f(g)<a$. Then, it turns to cases 1, 3, 4, 5.

Case 7. $d<\epsilon _{0}<c$. Since $f^{\prime }(x)>0$ when $d<x<c$, $%
0<\epsilon _{1}<a$. Then, it turns to cases 1, 3, 4, 5.

(B). When $\epsilon _{0}\in (c,1)$, the deviation from the target state
converges to the fixed-points $a$ or $0$.

When $\epsilon _{0}>c$, by property (1.2) the sequence $\{\epsilon _{i}\}$
decreases monotonically from $\epsilon _{0}$ to $\epsilon _{i^{\ast }\text{ }%
}\leq c$. Case 1, if $\epsilon _{i^{\ast }\text{ }}=d$, then $\epsilon
_{i}=0 $ for any $i>i^{\ast }$.\ Case 2. Otherwise, by the above (A) $%
\lim_{m\rightarrow \infty }\epsilon _{m}=a$.

Example 3. For the Phase-$2\pi /3$ search, $a=1/3$. The iteration equation
becomes $\epsilon _{m}=\epsilon _{m-1}(3\epsilon _{m-1}-2)^{2}$. Let $%
\epsilon _{0}=0.99999$. We have the following iterations. See Fig. 1.

$\epsilon _{1}=0.999\,93$, $\ \epsilon _{2}=0.999\,51$, \ \ $\epsilon
_{3}=0.996\,57$, $\ \epsilon _{4}=0.976\,17$,

$\epsilon _{5}=0.841\,59$, $\ \epsilon _{6}=0.231\,76$, $\ \ \epsilon
_{7}=\allowbreak 0.394\,52$, $\ \epsilon _{8}=\allowbreak 0.263\,50$,

$\epsilon _{9}=\allowbreak 0.385\,47$, $\ \epsilon _{10}=\allowbreak
0.274\,32$, \ $\epsilon _{11}=\allowbreak 0.380\,05$, $\ \epsilon
_{12}=\allowbreak 0.280\,99$,

$\epsilon _{13}=\allowbreak 0.376\,17$.

\subsection{$2\protect\pi /3<\protect\theta \leq \protect\pi $, the Phase-$%
\protect\theta $ search does not converge.}

For the Phase-$\theta $ search, $b<r<a<d<c$. From Eq. (\ref{fp-eq1}), we
have the following property.

Property 5

(5.1). $a<\epsilon _{m}\leq g$ whenever $b<\epsilon _{m-1}<a$;

(5.2). $0\leq \epsilon _{m}<a$ whenever $a<\epsilon _{m-1}<c$ or $\epsilon
_{m-1}<b$.

For large $\epsilon $,\ by property (1.2), the sequence $\{\epsilon _{i}\}$
decreases monotonically from $\epsilon _{0}$ to $\epsilon _{i^{\ast }}(\leq
c)$. If $\epsilon _{i^{\ast }}=d$, then $\epsilon _{i}=0$ for any $i>i^{\ast
}$. If $\epsilon _{i^{\ast }}=a$, $b$, or $c$, then $\epsilon _{i}=a$ when $%
i>i^{\ast }$. Otherwise, when $i>i^{\ast }$, $\epsilon _{i}$ oscillate
around the fixed point $a$ by property 1. However, the sequence $\{\epsilon
_{i}\}$ does not converges because $a$, $0$ and $1$ are repulsive
fixed-points.

Example 4. For the Phase-$\pi $ search, the iteration equation becomes $%
\epsilon _{m}=\epsilon _{m-1}(4\epsilon _{m-1}-3)^{2}$, $a=1/2$. Let $%
\epsilon =0.99999$. We have the following iterations. See Fig. 1.

$\epsilon _{1}==0.999\,91$, $\ \ \epsilon _{2}=0.999\,19,$ $\ \epsilon
_{3}=0.992\,73$, $\ \epsilon _{4}=0.935\,83,$

$\epsilon _{5}=0.517\,07$, $\ \ \ \epsilon _{6}=0.448\,87$, \ \ $\epsilon
_{7}=0.651\,25$, $\ \epsilon _{8}=0.101\,61,$

$\epsilon _{9}=0.683\,49$, $\ \ \ \epsilon _{10}=4.\,837\,6\times 10^{-2}$.

Clearly, the sequence $\{\epsilon _{i}\}$ monotonically decreases from $%
\epsilon _{0}$ to $\epsilon _{6}$. Note that after the sixth iteration, $%
\epsilon _{m}$ oscillate around the fixed point $1/2$.

\section{A comparison of rates of convergence after any number of iterations}

For the Phase-$\pi /3$ search, let the iteration equation be $\epsilon
_{m}(\pi /3)=(\epsilon _{m-1}(\pi /3))^{3}$, where $\epsilon _{m}(\pi /3)$
is the failure probability of the Phase-$\pi /3$ search algorithm after $m$
iterations.\ For the Phase-$\theta $ search, we can rewrite Eq. (\ref%
{g-iteration}) as $\epsilon _{m}(\theta )=4(1-\cos \theta )^{2}\epsilon
_{m-1}(\theta )(\epsilon _{m-1}(\theta )-d)^{2}$, where the $\epsilon
_{m}(\theta )$\ is the failure probability of the Phase-$\theta $ search
algorithm after $m$ iterations.\ We want to compare the failure probability
of the Phase-$\theta $ ($\neq \pi /3$) search algorithm with the one of the
Phase-$\pi /3$ search after $m$ iterations.\ It is known that the less the
failure probability is, the faster the algorithm converges.\ By factoring,

\begin{eqnarray}
\epsilon _{m}(\theta )-\epsilon _{m}(\pi /3) &=&  \notag \\
&&\epsilon _{m-1}(\theta )(2\cos \theta -1)(1-\epsilon _{m-1}(\theta
))(3-2\cos \theta )\ast  \notag \\
&&(\epsilon _{m-1}(\theta )-\frac{1-2\cos \theta }{3-2\cos \theta }%
)+\epsilon _{m-1}^{3}(\theta )-\epsilon _{m-1}^{3}(\pi /3).  \label{compare}
\end{eqnarray}

We have the following results.

(1). $\pi /3<\theta \leq \pi $

Case 1. For large $\epsilon $, the Phase-$\theta $ search converges faster
than the Phase-$\pi /3$ search for $m$ iterations until $\epsilon
_{m-1}(\theta )<$ $\frac{1-2\cos \theta }{3-2\cos \theta }$.

In \cite{LDF07a}, we show if $\epsilon _{0}(\theta )=\epsilon _{0}(\pi
/3)=\epsilon >\frac{1-2\cos \theta }{3-2\cos \theta }$ then $\epsilon
_{1}(\theta )<\epsilon _{1}(\pi /3)=\epsilon ^{3}$. If $\epsilon
_{m-1}(\theta )>\frac{1-2\cos \theta }{3-2\cos \theta }$ and $\epsilon
_{m-1}(\theta )<\epsilon _{m-1}(\pi /3)$, then by Eq. (\ref{compare})$\
\epsilon _{m}(\theta )<\epsilon _{m}(\pi /3)$. Thus, $\epsilon _{i}(\theta
)<\epsilon _{i}(\pi /3)$, where $i=1$, $2$, ..., $m-1$, until $\epsilon
_{m-1}(\theta )<$ $\frac{1-2\cos \theta }{3-2\cos \theta }$. It says that
after $m$ iterations, the failure probability of\ the Phase-$\theta $ search
is less than the one of the Phase-$\pi /3$ search until $\epsilon
_{m-1}(\theta )<$ $\frac{1-2\cos \theta }{3-2\cos \theta }$. It suggests us
first to use the fixed-point search with large phase shifts for the large
size of database.

Case 2. For small $\epsilon $, the Phase-$\pi /3$ search converges faster
than the Phase-$\theta $ search for $m$ iterations until $\epsilon
_{m-1}(\theta )>\frac{1-2\cos \theta }{3-2\cos \theta }$.

\ In \cite{LDF07a}, we show if $\epsilon _{0}(\theta )=\epsilon _{0}(\pi
/3)=\epsilon <\frac{1-2\cos \theta }{3-2\cos \theta }$ then $\epsilon
_{1}(\theta )>\epsilon _{1}(\pi /3)=\epsilon ^{3}$. If $\epsilon
_{m-1}(\theta )<\frac{1-2\cos \theta }{3-2\cos \theta }$ and $\epsilon
_{m-1}(\theta )>\epsilon _{m-1}(\pi /3)$, then by Eq. (\ref{compare})$\
\epsilon _{m}(\theta )>\epsilon _{m}(\pi /3)$.

(2). When $0<\theta <\pi /3$, the Phase-$\pi /3$ search converges faster
than the Phase-$\theta $ search for any $\epsilon $ for any number of
iterations.

When $\epsilon _{0}(\theta )=\epsilon _{0}(\pi /3)=\epsilon $, in \cite%
{LDF07a} we show $\epsilon _{1}(\theta )>\epsilon _{1}(\pi /3)$. Assume that
$\epsilon _{m-1}(\theta )>\epsilon _{m-1}(\pi /3)$. From Eq. (\ref{compare}%
), it is easy to see that $\epsilon _{m}(\theta )>\epsilon _{m}(\pi /3)$.
Therefore, $\epsilon _{m}(\theta )>\epsilon _{m}(\pi /3)$ for any $m$.
Hence, when $0<\theta <\pi /3$,\ the Phase-$\pi /3$ search converges faster
than the Phase-$\theta $ search for any $\epsilon $\ for any number of
iterations.

\section{For any known $\protect\epsilon $, $O(n)$ iterations can find the
target state.}

Assume that a database has $N=2^{n}$ states (items). Then a state (an item)
is found with the probability of $1/N$\cite{Grover98}. In other words, the
failure probability $\epsilon =1-1/N$. It is known that the Phase-$\pi /3$
search converges the target state.\ In this section, we investigate how to
use the fixed-point search to find the target state in a database when $%
\epsilon $ is known. As discussed in \cite{Grover05}, the fixed-point search
is a recursive algorithm, therefore the number of queries grows
exponentially with the number of recursion levels. For example, the Phase-$%
\pi /3$ search at $i$-level recursion involves $q_{i}=(3^{i}-1)/2$ queries
\cite{Tulsi}.\ This implies that $O(n)$ iterations of the Phase-$\theta $
search involve the exponential queries.

\subsection{When $\protect\epsilon \leq 3/4$, only one iteration is needed
to find the target state.}

When $0\leq \epsilon \leq \frac{3}{4}$, $\left\vert 1-\frac{1}{2(1-\epsilon )%
}\right\vert \leq 1$. Let $\cos \theta =1-\frac{1}{2(1-\epsilon )}$. Then $%
D(\theta )=0$. Therefore, if $\epsilon $ is fixed and $0\leq \epsilon \leq
\frac{3}{4}$, then we choose $\theta =\arccos [1-\frac{1}{2(1-\epsilon )}],$
which is in $(\pi /3$ ,$\pi ]$, as phase shifts. The Phase-$\arccos [1-\frac{%
1}{2(1-\epsilon )}]$ search will obviously make the deviation vanish. It
means that one iteration will reach $t$ state if the $\theta $ is chosen as
phase shifts. Ref. \cite{LDF07a}.

\subsection{When $\protect\epsilon >3/4$, $O(n)$ iterations can find the
target state.}

\subsubsection{First use the Phase-$\protect\pi /3$ search}

For the Phase-$\pi /3$ search, $\epsilon _{n}=\epsilon ^{3^{n}}$. There
exists the least natural number $n^{\ast }$ such that $\epsilon ^{3^{n^{\ast
}}}\leq 3/4$. By calculating, $n^{\ast }=\lceil (\ln \ln \frac{4}{3}-\ln \ln
\frac{1}{\epsilon })/\ln 3\rceil $.

Lemma 1. For the Phase-$\pi /3$ search, $n^{\ast }=$ $O(n)$.

Proof. In the case of database search, Let $N=2^{n}$. Then $\epsilon
=1-2^{-n}$, and $\lim_{n\rightarrow +\infty }\frac{n^{\ast }}{n}=\frac{\ln 2%
}{\ln 3}$. Almost $\frac{n\ln 2}{\ln 3}\approx \allowbreak \lceil
0.63n\rceil $. Let $N=10^{n}$. Then $\epsilon =1-10^{-n}$, and $%
\lim_{n\rightarrow +\infty }\frac{n^{\ast }}{n}=\frac{\ln 10}{\ln 3}=\frac{1%
}{\lg 3}$. Almost $\frac{n}{\lg 3}\approx 2n$. Thus, $n^{\ast }=O(n)$.
Hence, when $\epsilon >3/4$, after $n^{\ast }$ iterations of the Phase-$\pi
/3$ search the failure probability $\epsilon _{n^{\ast }}\leq $ $3/4$. Then,
after one iteration of the Phase-$\arccos [1-\frac{1}{2(1-\epsilon _{n^{\ast
}})}]$ search by using the result in section 4.1, it will reach $t$ state.

Example 5. Let $N=10^{4}$. Then $\epsilon =1-10^{-4}$, $n^{\ast }=8$,$\
\epsilon _{7}=\allowbreak 0.803\,32$, \ $\epsilon _{8}=\allowbreak 0.518\,4$%
. See Fig.1. However, for this purpose, it only needs 4 iterations for the
Phase-$\pi $ search. See example 7.

Example 6. Let $N=2^{10}$. Then $\epsilon =1-2^{-10}$, $n^{\ast }=6$, $%
\epsilon _{5}=\allowbreak 0.788\,56$, $\epsilon _{6}=\allowbreak 0.490\,35$.

\subsubsection{First use the Phase-$\protect\theta $ ($\neq \protect\pi /3$)
search}

Let $\epsilon >3/4$. Then, by property (1.2), for the Phase-$\theta $
search, there exists the least natural number $m^{\ast }(\theta )$ such that
$\ \epsilon _{0}>\epsilon _{1}>...>\epsilon _{m^{\ast }(\theta
)-1}(>3/4)>\epsilon _{m^{\ast }(\theta )}(\leq 3/4)$.... Thus, after $%
m^{\ast }(\theta )$ iterations of the Phase-$\theta $ search, the failure
probability $\epsilon _{m^{\ast }(\theta )}\leq 3/4$. Then, after one
iteration for the Phase-$\arccos [1-\frac{1}{2(1-\epsilon _{m(\theta )^{\ast
}})}]$ search by using the result in section 4.1, it will reach $t$ state.

Next let us calculate $m^{\ast }(\theta )$.\ Let $\delta =1-\epsilon $,
where $\delta $ is the success probability. When $\epsilon $ is close to $1$%
, $\delta $ is close to $0$.\ Then, for large $\epsilon $, by induction $%
\epsilon _{l}=1-[1+4(1-\cos \theta )]^{l}\delta +O(\delta ^{2})$. Thus, $%
\epsilon _{l}\approx 1-[1+4(1-\cos \theta )]^{l}\delta $. By this
approximate formula of $\epsilon _{l}$, $m^{\ast }(\theta )$ $\approx
M^{\ast }(\theta )=\lceil \frac{-2\lg 2-\lg \delta }{\lg (1+4(1-\cos \theta
))}\rceil $.

In the case of database search, let $N=2^{n}$. Then $\epsilon =1-2^{-n}$, $%
\delta =2^{-n}$, and $m^{\ast }(\theta )$ $\approx $ $M^{\ast }(\theta
)=\lceil \frac{(n-2)\lg 2}{\lg (1+4(1-\cos \theta ))}\rceil $. For the Phase-%
$\pi /3$ search, $M^{\ast }(\pi /3)=\lceil \frac{n\ln 2}{\ln 3}-\frac{2\ln 2%
}{\ln 3}\rceil $. Note that $\frac{2\ln 2}{\ln 3}=\allowbreak
1.\,\allowbreak 261\,9$. Therefore, when $n$ is large enough $M^{\ast }(\pi
/3)\approx $ $m^{\ast }(\pi /3)=n^{\ast }$. For the Phase-$\pi $ search, $%
m^{\ast }(\pi )\approx M^{\ast }(\pi )=\lceil (n-2)\lg 2/(2\lg 3)\rceil
\approx (\lg 2)n$. See Table (I).

Let $N=10^{n}$.\ Then $\epsilon =1-10^{-n}$, $\delta =10^{-n}$, and $m^{\ast
}(\theta )\approx M^{\ast }(\theta )=\lceil \frac{n-2\lg 2}{\lg (1+4(1-\cos
\theta ))}\rceil $. For the Phase-$\pi /3$ search, $M^{\ast }(\pi /3)=\lceil
\frac{n}{\lg 3}-\frac{2\lg 2}{\lg 3}\rceil $. Note that $\frac{2\lg 2}{\lg 3}%
=\frac{2\ln 2}{\ln 3}$. Therefore, when $n$ is large enough $M^{\ast }(\pi
/3)\approx $ $m^{\ast }(\pi /3)=n^{\ast }$. For the Phase-$\pi $ search, $%
m^{\ast }(\pi )\approx M^{\ast }(\pi )=\lceil (n-2\lg 2)/(2\lg 3)\rceil
\approx n$. See

Example 7. Let $N=10^{4}$. Then $\epsilon =1-10^{-4}$,\ $M^{\ast }(\pi )=4$,
$\epsilon _{4}=0.475\,32$. See Table (II).

Lemma 2. For the Phase-$\theta $ $(\neq \pi /3)$ search, $m^{\ast }(\theta
)= $ $O(n)$.

Proof. When $\pi /3<\theta \leq \pi $, as discussed in case 1 of (1) in Sec.
3, $m^{\ast }(\theta )<n^{\ast }$. By lemma 1, this lemma holds. When $%
0<\theta <\pi /3$, from the approximate formula of $m^{\ast }(\theta )$, $%
m^{\ast }(\theta )=$ $O(n)$.

Remark. $M^{\ast }(\theta )$ monotonically decreases as $\theta $ increases
from $0$ to $\pi $, especially $\frac{M^{\ast }(\pi )}{n^{\ast }}\approx 1/2$%
. Therefore, we suggest first to use Phase-$\pi $ search for $m^{\ast }(\pi
) $ times to get the failure probability $\epsilon _{m^{\ast }}\leq 3/4$.

\section{Summary}

In this paper, we investigate convergence performance of the Phase-$\theta $
search\ for any number of iterations. We discuss the convergence region and
rate of the Phase-$\theta $ search and study the convergence behavior of the
Phase-$\theta $ search\ for different initial $\epsilon _{0}$.

\textbf{Acknowledgement}

We want to thank the reviewer of \cite{LDF07a} for suggesting us to study
the convergence behavior of the fixed-point search with general but equal
phase shifts\ for any number of iterations.

\section{Appendix A}

Proof. Since $f^{\prime }(x)<0$ when $r<x<d$, $f(g)\leq \epsilon _{1}<a$.
Note that $r<f(g)$. Thus, $r<f(g)\leq \epsilon _{1}<a$. Let $%
f^{(k)}(x)=f(f^{(k-1)}(x))$. Since $f^{\prime }(x)<0$, $a<\epsilon _{2}\leq
f^{(2)}(g)<f(r)=g$ and $r<f(g)<f^{(3)}(g)\leq \epsilon _{3}<a$. By
induction, generally $a<\epsilon _{2k}\leq
f^{(2k)}(g)<f^{(2k-2)}(g)<...f^{(2)}(g)<g$ and $%
r<f(g)<...<f^{(2k-1)}(g)<f^{(2k+1)}(g)\leq \epsilon _{2k+1}<a$. That is, $%
\epsilon _{i}$ oscillate around the fixed point $a$ by property 1 and
between $f^{(2k)}(g)$\ and $f^{(2k+1)}(g)$.\ It is plain that the sequence $%
\{f^{(2k)}(g)\}$ decreases monotonically as $k$ increases while the sequence
$\{f^{(2k+1)}(g)\}$ increases monotonically as $k$ does. Hence, the
sequences $\{f^{(2k)}(g)\}$ and $\{f^{(2k+1)}(g)\}$ have limits. Let $%
\lim_{k\rightarrow \infty }f^{(2k)}(g)=\alpha $ and $\lim_{k\rightarrow
\infty }f^{(2k+1)}(g)=\beta $. Clearly, $\alpha $, $\beta <d$. From Eq. (\ref%
{g-iteration}), $f^{(2k)}(g)=4(1-\cos \theta
)^{2})f^{(2k-1)}(g)(f^{(2k-1)}(g)-d)^{2}$ and $f^{(2k+1)}(g)=4(1-\cos \theta
)^{2})f^{(2k)}(g)(f^{(2k)}(g)-d)^{2}$. By taking the limits, we obtain $%
\alpha =4(1-\cos \theta )^{2})\beta (\beta -d)^{2}$ and $\beta =4(1-\cos
\theta )^{2})\alpha (\alpha -d)^{2}$. By substituting, $\beta =[4(1-\cos
\theta )^{2})]^{2}\beta (\beta -d)^{2}(\alpha -d)^{2}$. By cancelling, $%
[4(1-\cos \theta )^{2})]^{2}(\beta -d)^{2}(\alpha -d)^{2}=1$. Then, there
are two cases. Case 1. $4(1-\cos \theta )^{2})(d-\beta )(d-\alpha )=1$. By
solving this equation, $\alpha =\beta =1$ or $\alpha =\beta =a$. Since $%
\alpha $, $\beta <d<1$, then $\alpha =\beta =a$. Case 2. $4(1-\cos \theta
)^{2})(d-\beta )(d-\alpha )=-1$. There is no solution because $\alpha $, $%
\beta <d$. Therefore, $\lim_{k\rightarrow \infty
}f^{(2k)}(g)=\lim_{k\rightarrow \infty }f^{(2k+1)}(g)=a$. Then, $%
\lim_{k\rightarrow \infty }f^{(k)}(g)=a$, and also $\lim_{m\rightarrow
\infty }\epsilon _{m}=a$. We finish the proof.

\end{document}